\documentclass[a4paper]{jpconf}
\usepackage{graphicx}
\usepackage{amsmath}
\usepackage{amssymb}

\newcommand{\dd}{\partial}

\begin{document}
\title{3+1 dimensional viscous hydrodynamics at high baryon densities}

\author{Iu.A.~Karpenko$^{1,2}$, M. Bleicher$^{1,3}$, P. Huovinen$^{1}$ and H. Petersen$^{1}$}

\address{$^1$ Frankfurt Institute for Advanced Studies, Ruth-Moufang-Stra{\ss}e 1, 60438 Frankfurt am Main, Germany}
\address{$^2$ Bogolyubov Institute for Theoretical Physics, 14-b, Metrolohichna str., 03680 Kiev, Ukraine}
\address{$^3$ Institute for Theoretical Physics, Johann Wolfgang Goethe Universit\"at, Max-von-Laue-Str. 1, 60438 Frankfurt am Main, Germany}
\ead{karpenko@fias.uni-frankfurt.de}

\begin{abstract}
We apply a 3+1D viscous hydrodynamic + cascade model to the heavy ion collision reactions with $\sqrt{s_{NN}}=6.3\dots39$ GeV. To accommodate the model for a given collision energy range, the initial conditions for hydrodynamic phase are taken from UrQMD, and the equation of state at finite baryon density is based on Chiral model coupled to the Polyakov loop.

We study the collision energy dependence of pion and kaon rapidity distributions and $m_T$-spectra, as well as charged hadron elliptic flow and how shear viscosity affects them. The model calculations are compared to the data for Pb-Pb collisions at CERN SPS, as well as for Au-Au collisions in the Beam Energy Scan (BES) program energies at BNL RHIC. The data favours the value of shear viscosity $\eta/s\gtrsim0.2$ for this collision energy range.
\end{abstract}

\section{Introduction}
Hydrodynamic approach is well established for description of hot and dense matter created in ultrarelativistic heavy ion collisions at BNL Relativistic Heavy Ion Collider (RHIC) and CERN Large Hadron Collider (LHC). Its success is based on reproduction of bulk observables such as radial flow, elliptic and higher order flow harmonics \cite{huovinenReview}, femtoscopy \cite{Karp} etc.\ in so-called hybrid models coupling viscous (or inviscid) hydrodynamic phase to a hadron cascade.

In this paper we report on the application of the viscous+cascade model tuned to full RHIC energy to heavy ion collisions at lower collision energies. To accommodate the different collision energy regime, we change the equation of state (EoS) and the initial conditions, and keep all the other parameters fixed.

\section{Viscous hydro+cascade model}
The approximation of boost-invariant scaling flow is well-justified for hydrodynamic evolution of matter around midrapidity interval at top RHIC energies and above. However one should not pursue this assumption at lower energies, since at $\sqrt{s_{NN}}\approx 20$ GeV and below it becomes invalid because of the absence of plateau in rapidity distribution of produced particles.

\begin{figure}
\includegraphics[width=18pc]{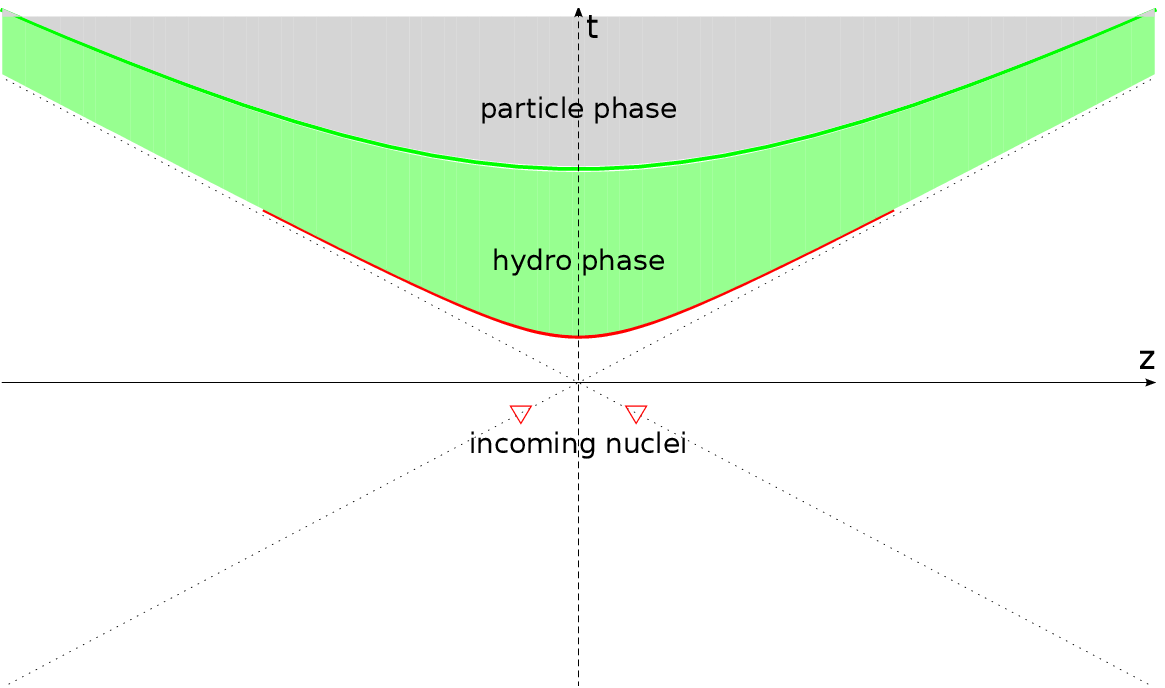}
\includegraphics[width=18pc]{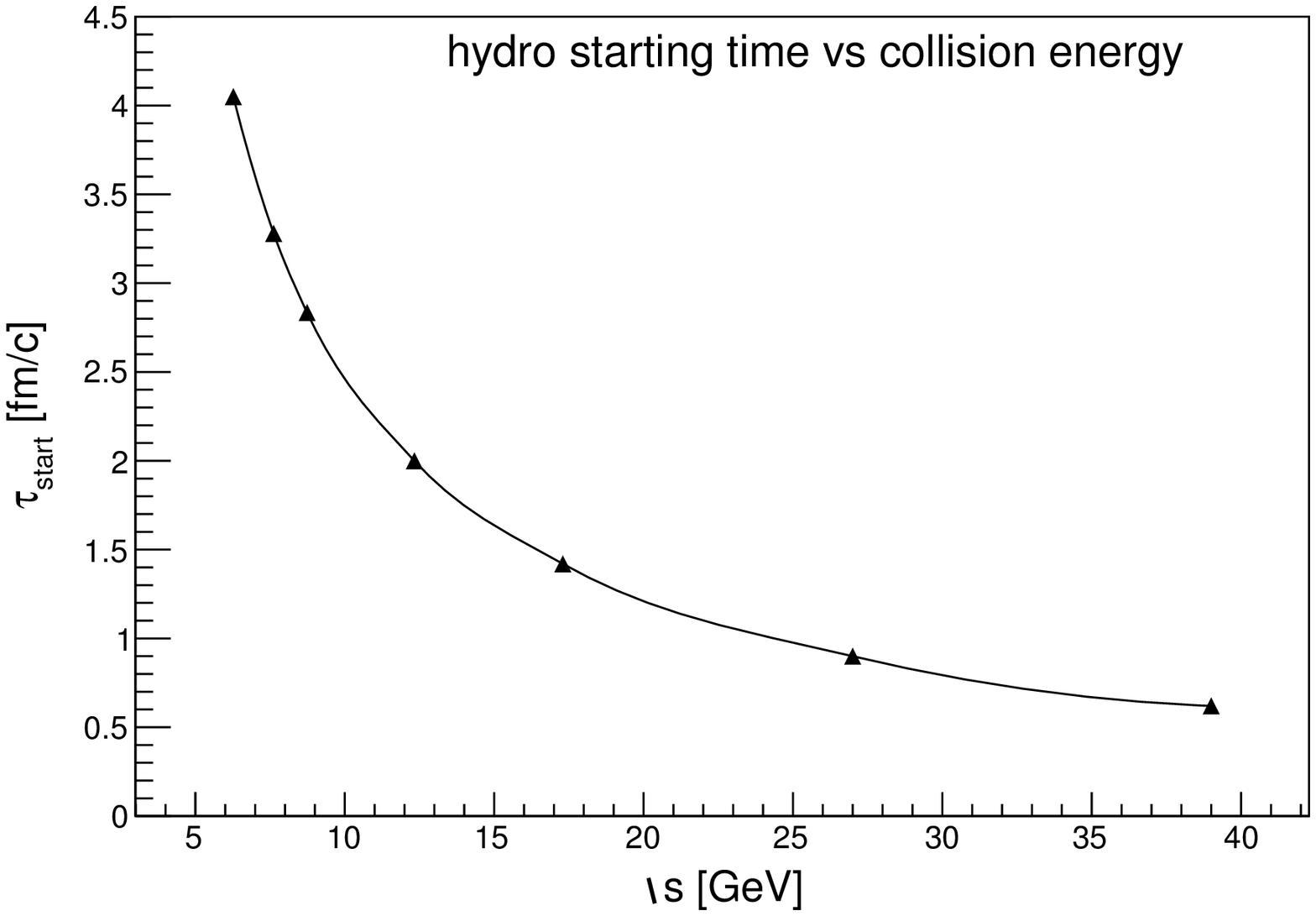}\\
\caption{\label{figEvolution} Left: Sketch of space-time picture of evolution, the different phases are marked by different colour. Right: starting time for hydrodynamic phase as a function of $\sqrt{s_{NN}}$ according to Eq. \ref{eqTau0}.}
\end{figure}

We avoid this issue by using rapidity-dependent initial conditions together with 3D hydrodynamic evolution. The model employed in the present studies consists of components, each corresponding to the certain stage of nucleus-nucleus collision. The space-time picture is sketched on Fig. \ref{figEvolution}.

\textbf{Pre-thermal phase.} Ultrarelativistic Quantum Molecular Dynamics (UrQMD) model \cite{urqmd} is used for the description of initial stage dynamics. The two nuclei are initialized according to Wood-Saxon distributions and the binary interactions are taken into account until a hypersurface at constant $\tau=\sqrt{t^2-z^2}$. We run UrQMD many times to get smooth averaged 3-dimensional distribution of particles which cross this hypersurface. The energy and momentum of particles is then converted to energy and momentum densities of the fluid. 
In addition to energy/momentum densities, initial baryon and charge densities are non-zero and obtained from UrQMD, evolved in the hydro stage and accounted for in the particlization procedure. On the other hand, net strangeness density is set to zero.

\textbf{Hydrodynamic phase.} We start the hydrodynamic phase at 
\begin{equation}
\tau_0=2R/\sqrt{(\sqrt{s}/2m_N)^2-1} \label{eqTau0}
\end{equation}
where $R$ is a radius of nucleus and $m_N$ is a nucleon mass. This corresponds to the time when the two nuclei have passed through each other. The values of starting time for different collision energies are shown on Fig. \ref{figEvolution}, right. The hydrodynamic equations are then solved in Milne ($\tau-\eta-r_x-r_y$) coordinates. 

Another important ingredient of the model relevant to lower collision energies is the equation of state (EoS) for finite baryon density. We employ the Chiral model based EoS \cite{ChiralEoS}, which features correct asymptotic degrees of freedom, i.e. quarks and gluons at high temperature and hadrons in the low-temperature limits, crossover-type transition between confined and deconfined matter for all values of $\mu_B$ and qualitatively agrees with lattice QCD data at $\mu_B=0$.

We employ Israel-Stewart framework for relativistic viscous hydrodynamics \cite{IsraelStewart}. Several variants of Israel-Stewart equations exist in literature, the differences between them coming from slightly different assumptions in the derivation \cite{denicol}. In particular we stick to the following choice for equations for the shear stress tensor:
\begin{equation}
\langle u^\gamma \dd_{;\gamma} \pi^{\mu\nu}\rangle=-\frac{\pi^{\mu\nu}-\pi_\text{NS}^{\mu\nu}}{\tau_\pi}-\frac 4 3 \pi^{\mu\nu}\dd_{;\gamma}u^\gamma
\end{equation}
where $\dd_{;\gamma}$ denotes covariant derivative and the brackets $\langle A^{\mu\nu}\rangle$ denote the symmetric, traceless and orthogonal to $u^\mu$ part of $A^{\mu\nu}$. For the purpose of the current study we consider only the effects of shear viscosity, fixing bulk viscosity to zero, $\zeta/s=0$. We do not include the baryon/electric charge diffusion either.
For viscous hydro simulations, we initialize the shear stress tensor to zero. The relaxation time for shear, $\tau_\pi$, is taken as $\tau_\pi=3\eta/(sT)$.

\textbf{Particlization and hadron corona.} Motivated by previous studies at $\sqrt{s_{NN}}=200$ A GeV RHIC and LHC energies \cite{Karp}, we fix the transition from fluid to particle description (so-called particlization) to take place at the constant energy density $\epsilon_\text{sw}=0.5$~GeV/fm$^3$, when the relevant degrees of freedom are hadrons. Cornelius subroutine \cite{Cornelius} is used to calculate the 3-volume elements $d\sigma_\mu$ of the transition hypersurface. It is important to note that while we use fixed energy density as transition criterion for all collision energies, the average net baryon density increases with decreasing collision energy. This corresponds to the increase of average temperature and decrease of baryon chemical potential with increasing collision energy, resembling the results for the collision energy dependence of chemical freezeout parameters from thermal model studies \cite{thermalModel}.
Since the EoS used has features (mean field, ) which somewhat shift the values of thermodynamic quantities from those in free hadron-resonance gas (HRG) approximation, we switch to free HRG EoS to sample the hadrons according to Cooper-Frye prescription. We recalculate the energy density, pressure, flow velocity and corresponding thermodynamical quantities from energy-momentum tensor using free HRG EoS, and employ them when sampling. To account for the effect of shear viscosity, we use the same corrections to the local equilibrium distribution functions for all hadron species:
\begin{equation}
 f_i(x,p)=f_{i,\text{eq}}(x,p)\left[1+(1\mp f_{i,\text{eq}})\frac{p_\mu p_\nu \pi^{\mu\nu}}{2T^2(\epsilon+p)}\right]
\end{equation}

Finally, UrQMD code is employed to calculate the further evolution of the hadron corona.

\section{Results}
We fix most of the parameters of the model as described above. The only free parameter left is the shear viscosity to entropy density ratio $\eta/s$ in hydro stage. Finally we perform simulations corresponding to various collision energy and $\eta/S$ values. Since a bulk of data from SPS exists for different observables ($dN/dy$, $p_T$-distributions, pion femtoscopy) one can first check how well the overall collective matter dynamics is described in the model. Thus we simulate Pb-Pb collisions at energies $E_\text{lab}=158, 80, 40, 30$ and $20$~A~GeV (i.e.\ $\sqrt{s_{NN}}=17.3,\dots,6.3$~GeV), and compare $m_T$ and $dN/dy$ distributions with the data from NA49 collaboration.

With decreasing beam energy, passage time calculated according to (\ref{eqTau0}) increases, whereas the average initial energy density for hydrodynamic expansion becomes smaller. The latter leads to shorter duration of hydro phase, provided that the criterion for fluid to particle transition is kept the same. Consequently, at the lowest SPS energy ($\sqrt{s_{NN}}=6.3$~GeV) the lifetimes of pre-hydro (4.1 fm/c) and hydro stages are comparable, and pre-hydrodynamic stage gives significant contribution to the development of transverse flow.

\begin{figure}[h]
 \includegraphics[width=18pc]{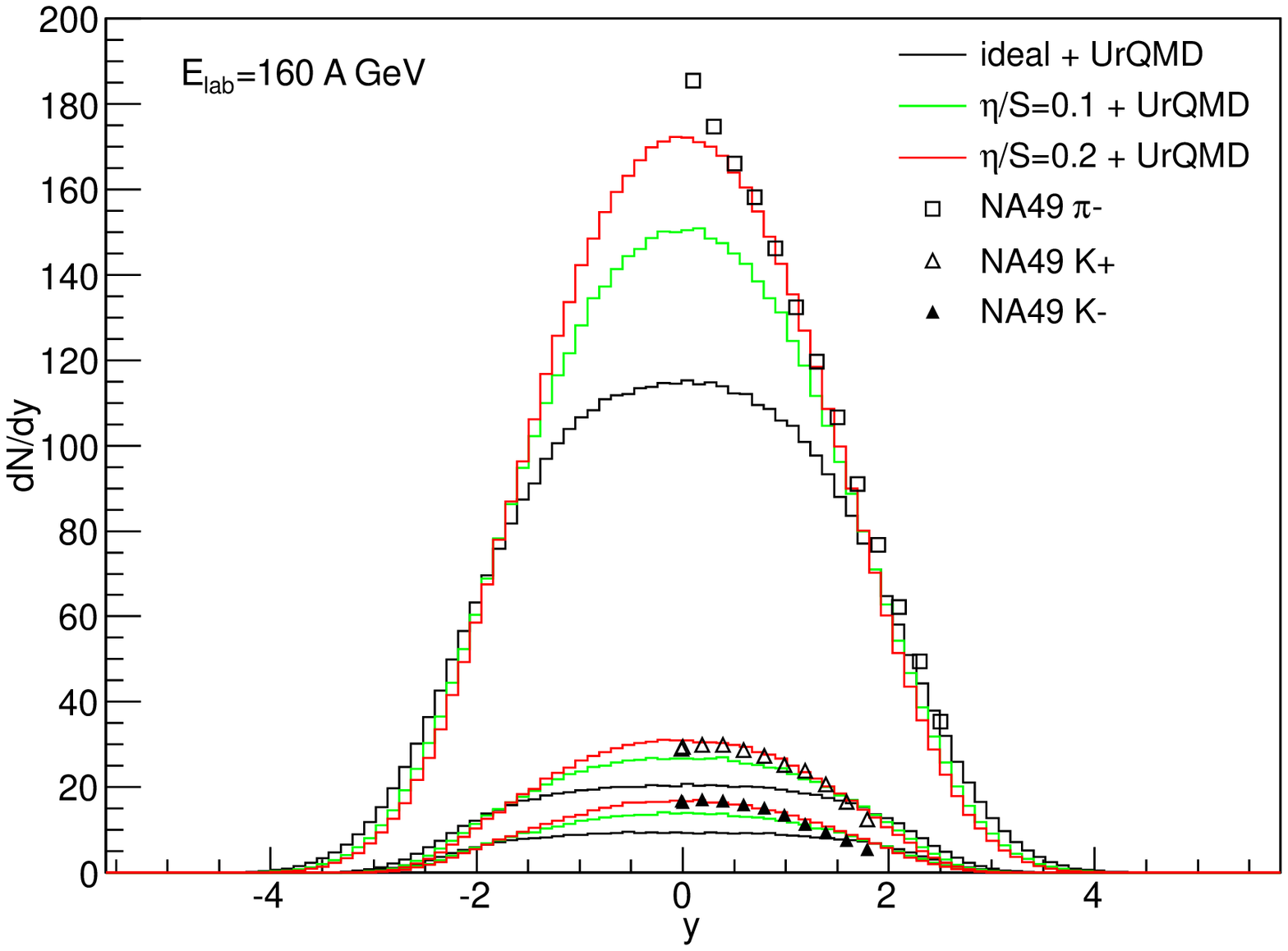}
 \includegraphics[width=17pc]{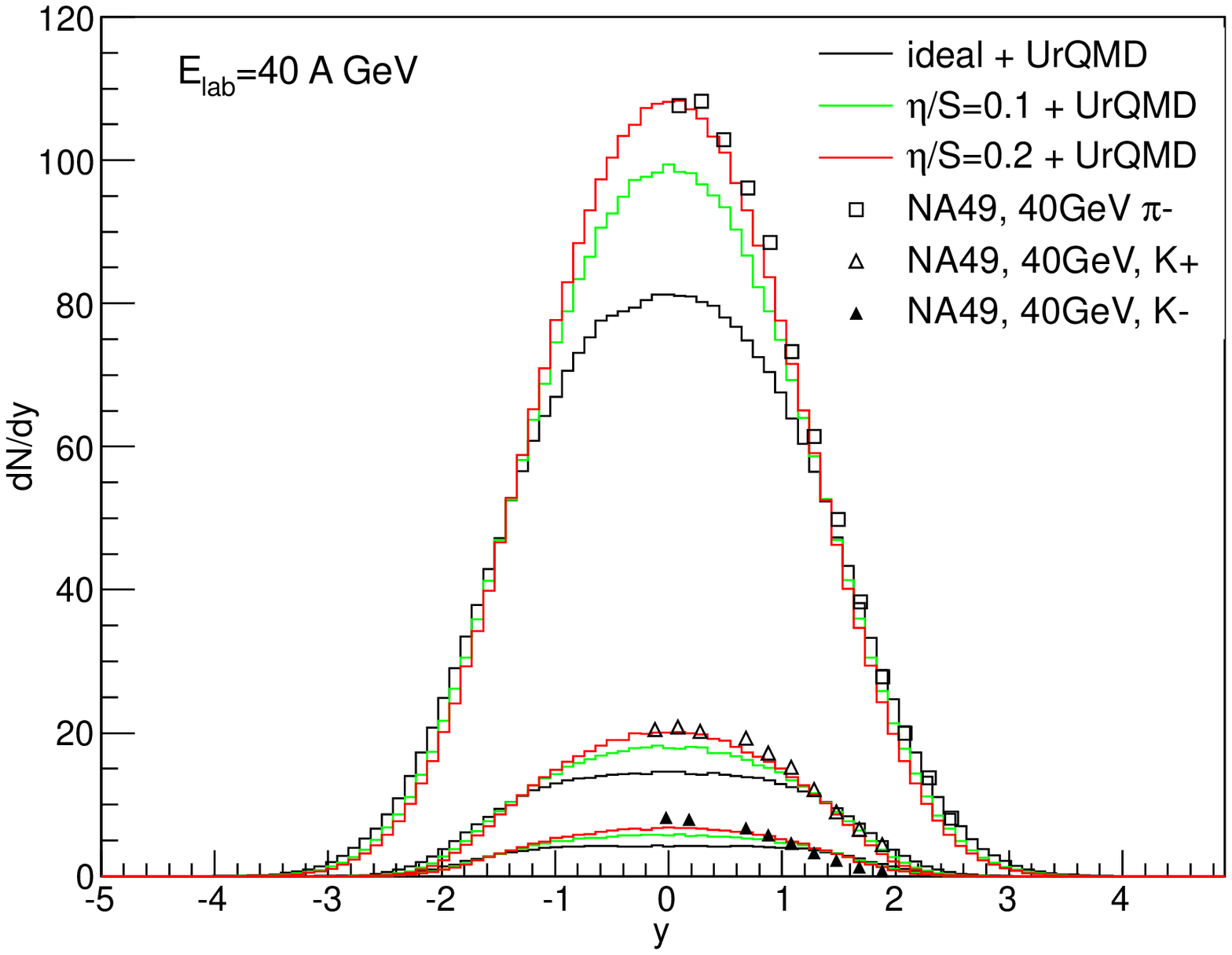}\\
 \caption{\label{figDnDy}Rapidity distributions for $\pi^-$,$K^+$ and $K^-$ for Pb-Pb collisions at $E_\text{lab}=158$ and $40$ A GeV (corresponding to $\sqrt{s_{NN}}=17.3$ and $6.3$ GeV). Model calculations are compared to NA49 data \cite{NA49-40-160}. The figures are taken from \cite{sqmProc}.}
\end{figure}

\begin{figure}[h]
 \includegraphics[width=18pc]{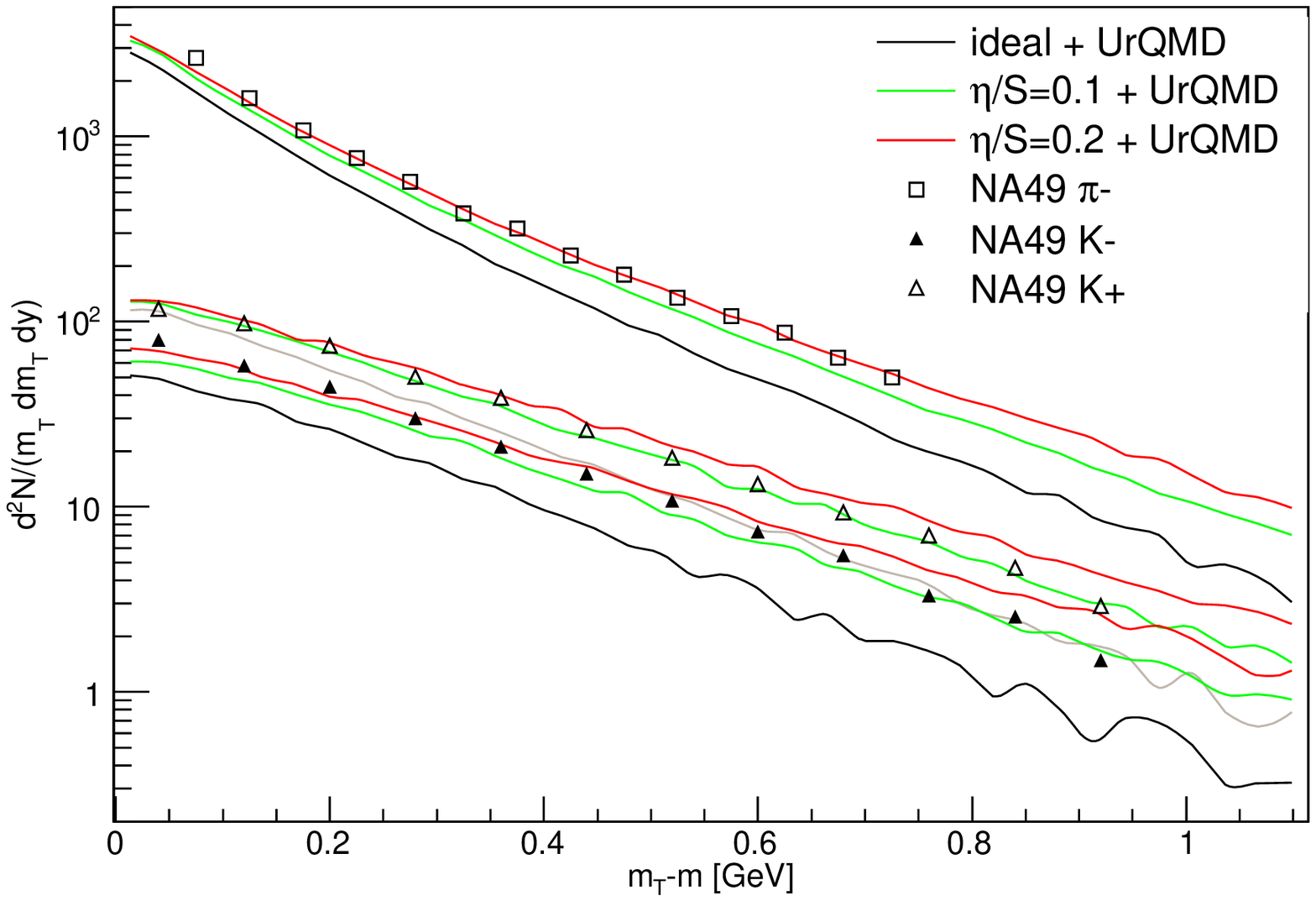}
 \includegraphics[width=18pc]{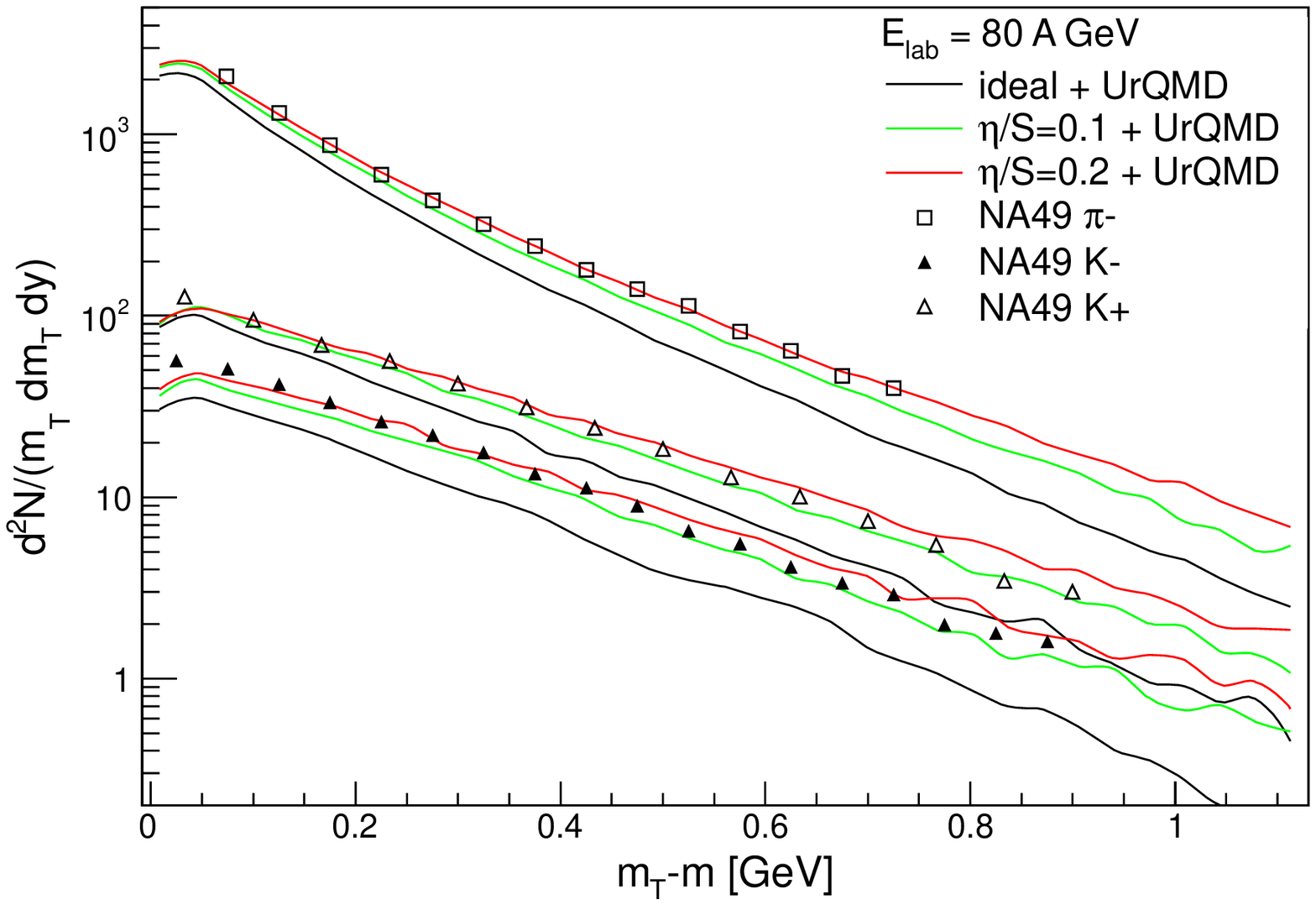}\\
 \includegraphics[width=18pc]{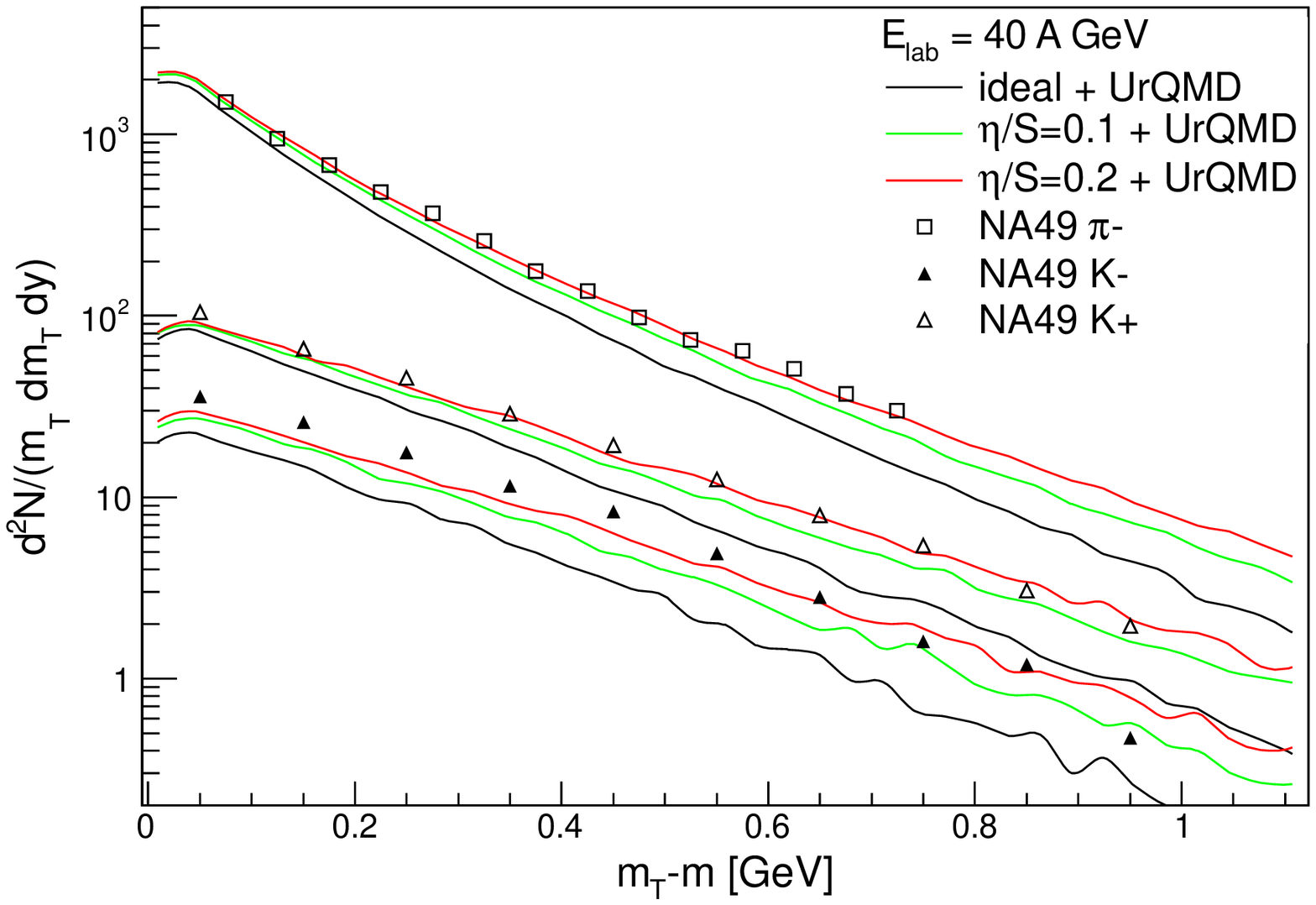}
 \includegraphics[width=18pc]{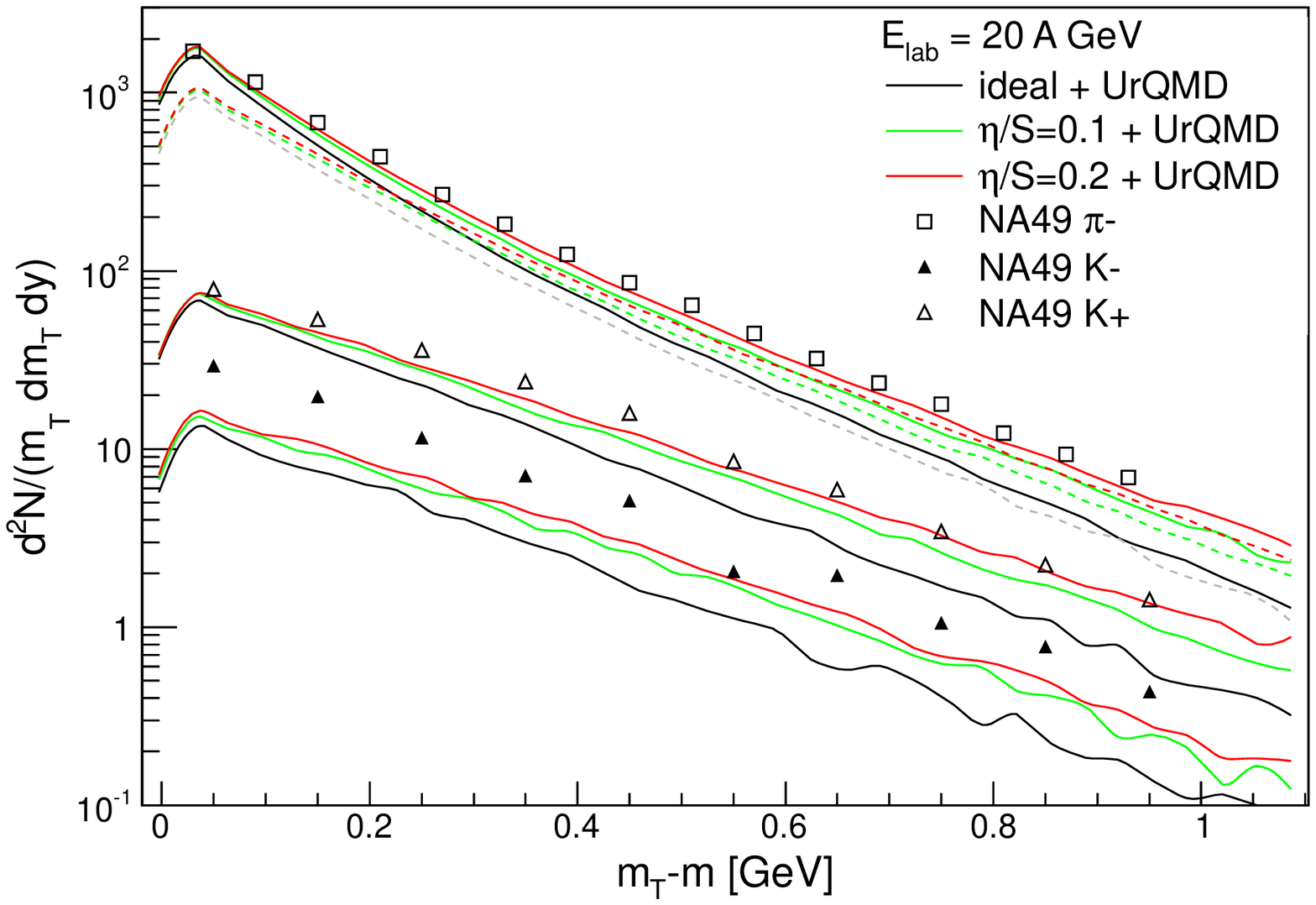}
 \caption{\label{figpt}$m_T$-spectra for $\pi^-$,$K^+$ and $K^-$ for Pb-Pb collisions at $E_\text{lab}=158, 80, 40, 20$~A~GeV ($\sqrt{s_{NN}}=17.3, 12.3, 8.8, 6.3$ GeV, respectively). Model calculations are compared to NA49 data \cite{NA49-20-30, NA49-40-160}. Dashed lines on bottom right plot: model calculations for $\pi^+$. The figures are taken from \cite{sqmProc}.}
\end{figure}

In the case of zero shear viscosity in hydro phase we underestimate the particle yields and radial flow at midrapidity.
However, the inclusion of shear viscosity in hydro phase increases the yields at midrapidity due to viscous entropy production, see Fig. \ref{figDnDy}. One can see that shear viscosity makes the $dN/dy$ profile narrower. The longitudinal expansion is weaker, and the expansion tends to be more spherical, as seen in the comparison of the $p_T$-spectra, Fig. \ref{figpt}: $p_T$-spectra become flatter, which is due to stronger transverse expansion and larger radial flow.

\begin{figure}
 \includegraphics[width=20pc]{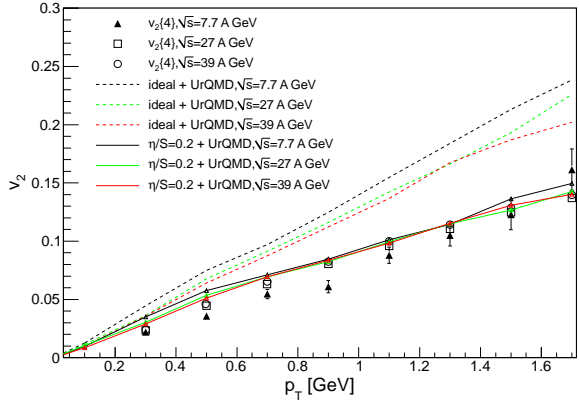}\hspace{2pc}
\begin{minipage}[b]{14pc}\caption{\label{figv2} $p_T$-differential elliptic flow of all charged hadrons for 20-30\% central Au-Au collisions at $\sqrt{s_{NN}}=7.7, 27$ and $39$~GeV. The data are from STAR collaboration \cite{starBESv2} The figures are taken from \cite{sqmProc}.}
\end{minipage}
\end{figure}

On Fig. \ref{figv2} we show $p_T$-differential charged hadron elliptic flow at collision energies $\sqrt{s_{NN}}=7.7, 27$ and $39$~A~GeV compared with the results from RHIC BES. Inviscid hydrodynamic phase leads to an overestimate of the data, while choosing $\eta/s=0.2$ greatly improves agreement with it. Here no attempt to fit the data is made; we only demonstrate the model results for $\eta/s=0, 0.1$ and $0.2$. So one can conclude that a consistent description of $v_2$, $dN/dy$, and $p_T$-spectra requires a value of $\eta/s$ which is somewhat larger than $0.2$, especially for lower energy points. The $\eta/s=0.2$ also makes $v_2(p_T)$ almost independent of the collision energy.

We conclude that the introduction of shear viscosity in the hybrid model consistently improves the description of the data in the low energy region. The suggested value of the effective shear viscosity is $\eta/s\ge0.2$ for both Pb-Pb collisions with $E_\text{lab}=20\dots158$ GeV at SPS and Au-Au collisions with $\sqrt{s}=7.7\dots39$ GeV in BES program at RHIC. For comparison, the typical $\eta/s$ value obtained from the fit to elliptic flow at $\sqrt{s_{NN}}=200$~GeV RHIC energy is 0.08 for the case of Monte Carlo Glauber initial state \cite{Heinz}.

\section*{Acknowledgements}
IK and HP acknowledge the financial support by the ExtreMe Matter Institute EMMI and Hessian LOEWE initiative. HP acknowledges funding by the Helmholtz Young Investigator Group VH-NG-822. Computational resources have been provided by the Center for Scientific Computing (CSC) at the Goethe-University of Frankfurt.

\end{document}